\documentclass{emulateapj}
\usepackage{xcolor}
\usepackage{amsmath}
\usepackage{graphicx}
\usepackage[titletoc]{appendix}
\newcommand{\degree}{$^{\circ}$}

\usepackage[unicode=true,pdfusetitle,bookmarks=true,bookmarksnumbered=false,bookmarksopen=true,breaklinks=true,pdfborder={0 0 0},backref=section,colorlinks=true]{hyperref}
\hypersetup{linkcolor=blue,citecolor=blue}

\defcitealias{casanova2017}{GL17}
\defcitealias{lazarian2018b}{LY18}
\defcitealias{yuen2017a}{YL17}
\defcitealias{lazarian1999}{LV99}
\defcitealias{goldreich1995}{GS95}

\makeatother
\shorttitle{Using starlight polarization to test magnetic field structure obtained with velocity gradients}
\shortauthors{Gonz\'alez-Casanova \& Lazarian}

\begin{document}
\title{Mapping of the structure of the galactic magnetic field with velocity gradients: test using star light polarization}
\author{Diego F. Gonz\'alez-Casanova and A. Lazarian}
\email{casanova, lazarian @ astro.wisc.edu}
\affil{Astronomy Department, University of Wisconsin-Madison, 475 North Charter Street, Madison, WI 53706-1582, USA}

\begin{abstract} 
We apply Velocity Channel Gradients (VChGs) and Reduced Velocity Centroids Gradients (RVCGs) to H{\sc i} data (the LAB survey), obtaining the plane-of-sky component (POS) of the magnetic field as a function of the relative velocity. Assuming a galactic rotation curve, we transformed the relative velocities to distances, constructing the first map of the POS magnetic field at every point of the celestial sphere as a function of the distance. To test the accuracy of our 3D distribution, we used a set of stars with known distances from the stellar polarization catalog. We compared the polarization directions that we obtain with the VChGs and RVCGs against the starlight polarization directions. We find a good correspondence between the derived magnetic field and measured polarization directions, which testifies to the accuracy of this new way of probing the 3D galactic POS magnetic field structure.
\end{abstract}

\section{Introduction}

Magnetic fields and turbulence are ubiquitous in the interstellar medium (ISM), acting as important drivers of galactic evolution and feedback mechanisms. In the ISM, they significantly affect key astrophysical phenomena such as star formation, cosmic ray acceleration and propagation, and feedback mechanisms \citep{elmegreen2004, mckee2007, crutcher2012, naab2016}. Their effects are imprinted in electron density fluctuations, aniso- tropies of spectroscopic data, and Doppler broadening, to name a few situations \citep{armstrong1995, lazarian2000,chepurnov2010}. It is therefore fundamental to understand the strength and direction of the magnetic field in the ISM, as well as its morphology across the entire Galaxy.

Zeeman splitting \citep[see][]{crutcher2010} provides the most detailed information about the line-of-sight (LOS) magnetic field in the ISM. The measurements are very time- consuming, and only upper limits of the magnetic fields are obtained in most cases. Far-infrared polarimetry of aligned dust \citep[see][]{andersson2015} provides 2D maps of the plane-of-sky (POS) magnetic fields directions. {\it Planck} data \citep{Planck2016a, Planck2016b} provided a revolution in our understanding of POS magnetic fields. However, to increase the resolution of the maps, one has to use either balloon suborbital instruments, e.g., BLASTPOL \citep[see][]{fissel2016} or wait for future space missions. By measuring the polarization from stars with known distances \citep[see][]{heiles2000}, one obtains insight into the 3D structure of the POS galactic magnetic field. This method of magnetic field sampling is limited, as it is possible only to sample magnetic fields in the direction toward the stars with known distances. In addition, all dust polarimetry techniques\footnote{Intrinsic instrumental polarization is another complication for polarimetry measurements.} suffer from existing uncertainties related to grain alignment and the failure of grains to align at high optical depths; \citep[see][for a review]{lazarian2015}.

In this paper, we discuss a new technique to obtain 3D maps of the POS magnetic field in atomic hydrogen that is based on the present-day understanding of MHD turbulence. The technique employs both velocity gradients and galactic rotation curves to map the POS distribution of magnetic field directions.

The first paper discussing the velocity gradients as a means of magnetic field tracing was \citet[][henceforth referred to as GL17]{casanova2017}. For this technique, velocity centroids \citep[see][]{esquivel2005} were employed to represent the velocity information. Given that galactic rotation causes gas to move at different velocities, we sought to use the velocity resolution intrinsic to radio observations to distinguish different parts of the interstellar gas. For this, \citet[][henceforth referred to as LY18]{lazarian2018b} introduced Velocity Channel Gradients (VChGs) and Reduced Velocity Centroids Gradients (RVCGs). LY18 proposed to use both tools to find the distribution of POS magnetic fields in 3D by employing the galactic rotation curve. Our paper is the first practical attempt to implement this idea. For our calculations, we use the procedure of sub-block averaging introduced in \citet{yuen2017a}. The latter procedure allows us to reliably trace the gradient directions and evaluate the uncertainties of our magnetic field direction measurements.

Our tracing of the field directions in H{\sc i} has some similarities to the procedure of filament identification in \citet{clark2014, clark2015}. There, it was found empirically that, for high galactic latitude H{\sc i}, filamentary structures can be identified in velocity channel maps. These filaments tend to align with the magnetic field as measured by Planck polarimetry. We believe that the measured filaments are mostly velocity caustics rather than the real density structures. This follows from the theory of the statistics of fluctuations in the Position-Position-Velocity (PPV) space formulated in \citet[][]{lazarian2000} \cite[see also][for a review]{lazarian2009}. In our view, the observed ``filaments" and the gradients in velocity channel maps that we employ within the VChG technique both measure the proper- ties of turbulent velocities. We shall provide a detailed comparison of the two methods of magnetic field tracing elsewhere. We also note that, for denser regions (e.g.,molecular clouds), the filamentary structure in channel maps fades away and becomes unmeasurable, but velocity gradients can still trace these magnetic fields reliably (see LY18).

Our approach should not be confused with using density gradients within the Histograms of Relative Orientation technique (HRO) \citep{soler2013,soler2017}. The HRO is designed to compare the direction of density gradients and magnetic field as a function of column density. It requires polarization measurements and is not intended for obtaining the magnetic field direction on its own.

The VChGs and RVCGs that we employ here are particular incarnations of the Velocity Gradient Technique (VGT), which is based on a current understanding of MHD turbulence theory. The core paper is that on the theory of in-compressible MHD turbulence by \citet[][henceforth referred to as GS95]{goldreich1995}. This paper predicts the anisotropy of MHD turbulence. The theory of turbulent reconnection in \citet[][henceforth referred to as LV99]{lazarian1999} explains that eddies are not constrained from rotating perpendicular to the direction of the magnetic field that surrounds them. Magnetic tension is the force that resists any rotation that is not aligned with the magnetic field. As a result, the random turbulent driving preferentially induces motions perpendicular to the magnetic field that surrounds the eddies. Incidentally, this way of thinking naturally introduces the importance of local magnetic fields with respect to the motions of Alfvénic turbulence. The concept of a local system of reference was not a part of the original GS95 model of turbulence. However, numerical simulations \citep{cho2000,maron2001} have confirmed that the local system of reference is essential. We stress that, for the VGT, it is essential that turbulent eddies trace the local magnetic field around eddies rather than the mean magnetic field.\footnote{All earlier MHD turbulence theories, including the original formulation of the GS95 theory were done with respect to the mean field. The GS95 relations, however, are valid only in the local system of reference.} 

It is natural to expect the Alfv\'enic eddies that are not constrained by magnetic tension to create a Kolmogorov cascade with velocities $u_l \sim l^{1/3}_\perp$, where $l_{\bot}$ is measured with respect to the local direction of magnetic field. It is also evident that the eddies mixing magnetic field lines perpendicular to their direction should induce Alfv\'enic waves (velocity $V_A$) along the magnetic field. The equality of the eddy turnover time $l_{\bot}/u_l$ and the period of the Alfv{\'e}n wave $l_{\|}/V_A$ that is being induced by the eddy corresponds to the ``critical balance'' that can be written as

\begin{equation}
\label{chap:mapping-eq:cbe}
l^{-1}_\parallel V_A \approx  l^{-1}_\perp u_l    \;.
\end{equation}

It is easy to see that the predicted anisotropy of the eddies will be increasing as the eddy scale decreases. Indeed, $u_l$ decreases with the scale as $l^{1/3}$, which induces the change of the $l_{\bot}$ to $l_{\|}$ ratio.

Note the GS95 assumes the velocity at the injection scale to be $V_A$. This is the case of trans-Alfv\'enic turbulence. The anisotropy is even more prominent if the injection velocity is less than VA, as it is quantified in LV99.

Based on the velocity scaling ($u_l \sim l_\perp^{1/3}$), there are two takeaways. First, a gradient of the velocity increases as the eddy scale decreases ($\sim l^{-2/3}_\perp$). Second, the gradient is perpendicular to the direction of the local magnetic field. It is also possible to prove that the velocity gradients arising from the smallest eddies resolved by a telescope dominate the gradient signal despite LOS averaging \citep[see][]{lazarian2018}.

In this work, we obtain the VChG and RVCG from H{\sc i} spectroscopic data to acquire the structure of the POS galactic magnetic field as a function of relative velocity. The VChG and RVCG measurements are compared to stellar polarization measurements to determine the effectiveness of the technique. To compare the two measurements, we use the galactic rotation curve to relate the distances to velocities, allowing a direct comparison (at least in the disk of the Galaxy). We then derive a map of the structure of the perpendicular component of the galactic magnetic field as a function of distance.
In Section~\ref{chap:mapping-sec:data}, we describe the different surveys used; in Section~\ref{chap:mapping-sec:method}, we describe the method used; in Section~\ref{chap:mapping-sec:results}, we present the VChG and RVCG analysis on the H{\sc i} data and the alignment with stellar polarization; in Section~\ref{chap:mapping-sec:maps}, we present the structure of the galactic POS magnetic field; in Section~\ref{chap:mapping-sec:reqandlim}, we discuss the limits and requirements of the approach; in Section~\ref{chap:mapping-sec:discussion}, we discuss the impacts of the magnetic field map; and in Section~\ref{chap:mapping-sec:conclusion}, we give our conclusions.

\section{Observational Data}\label{chap:mapping-sec:data}

\subsection{H{\footnotesize I} data}\label{chap:mapping-subsec:hi}

The H{\sc i} data used comes from the Leiden/Argentine/Bonn (LAB) Survey \citep{kalberla2005}. The LAB survey is a combination of two independent H{\sc i} surveys, that together represent the first H{\sc i} map of the entire sky.  The LAB Survey is the most sensitive full sky survey kinematically, merging the Leiden/Dwingeloo survey ($\delta \ge$ -30\degree) \citep{hartmann1997} with the Instituto Argentino de Radioastronom\'{i}a Survey \citep{arnal2000, bajaja2005} ($\delta \le$ /-25\degree).  The spectral resolution is 1.3~km~s$^{-1}$ and the velocity coverage is 850~km~s$^{-1}$ (-450~km~s$^{-1}$~$<v_{LSR}<$~400~km~s$^{-1}$) with an angular resolution of 30' and 35.7' for the southern and northern parts respectively.

\subsection{Star polarization data}\label{chap:mapping-subsec:starpolarization}

The measures of starlight polarization come from the \citet{heiles2000} catalog, a compilation of 9286 stars. This catalog attempts to eliminate errors, provide positions, weight multiple observations of the same star, and provide distances to the stars. The stars in the catalog are located across the whole sky, but primarily in the plane of the Galaxy.

We only use polarization data from stars for which distances were measured with {\it Hipparcos} and found to be less than 4 kpc away. The 4 kpc radius limit represents the maximum distance {\it Hipparcos} can measure \citep{hipparcos1997,heiles2000}. This increases the accuracy of the distance measurements, enabling us to better compare the different observations. We further constrain the sample to only those stars where the polarization percentage is $pp$ $\geq$0.9, as defined in \citet{heiles2000} below:

\begin{equation}
\label{chap:mapping-eq:centroid}
pp = \sqrt{\langle Q \rangle ^2 + \langle U \rangle ^2}  \;,
\end{equation}
where $\langle Q \rangle$ and $\langle U \rangle$ are the Stokes parameters for each star.  We use the polarization percentage of 0.9 to increase the alignment between the star polarization and the inferred magnetic field. With the polarization and distance constraints there are 1139 stars in the sample.

We further constrain the sample by removing any stars 10\degree from the galactic center or anticenter. This cut prevents ambiguities related to the measurement of relative velocities. This is particularly important because we will transform from velocities to distances. With this final cut to the stellar catalog, we have 1043 stars with polarization information.

\section{Method} \label{chap:mapping-sec:method}

Using the spectroscopic H{\sc i} data, we constructed velocity channel maps and used two different approaches to the VGT to calculate the velocity gradients, namely the VChG and RVGC. To obtain the VChGs, the channel maps are formed with thin channels, in the way defined by \citet{lazarian2000}, because intensity fluctuations within thin channels represent velocity fluctuations rather than those induced by turbulent density. This procedure calculates the gradient at each of the channels using the sub-block averaging procedure in YL17. This is the way that we obtain the plane-of-sky (POS) magnetic field at each channel.

The second approach used is the RVCGs. The gradients are obtained in thicker velocity channels than the VChG, but for the velocity centroid, $S$, rather than the intensity. The RVCG covers the full velocity range of the data, and depends on the LOS velocity (just like the VChG). The RVCG and the VChG were found to have the same type of alignments between our inferred magnetic field direction and the direction from polarization measurements. Therefore, for the rest of the paper, we will only show results from the RVCG unless otherwise noted.

\begin{figure*}
\centering\includegraphics[width=0.8\linewidth,clip=true]{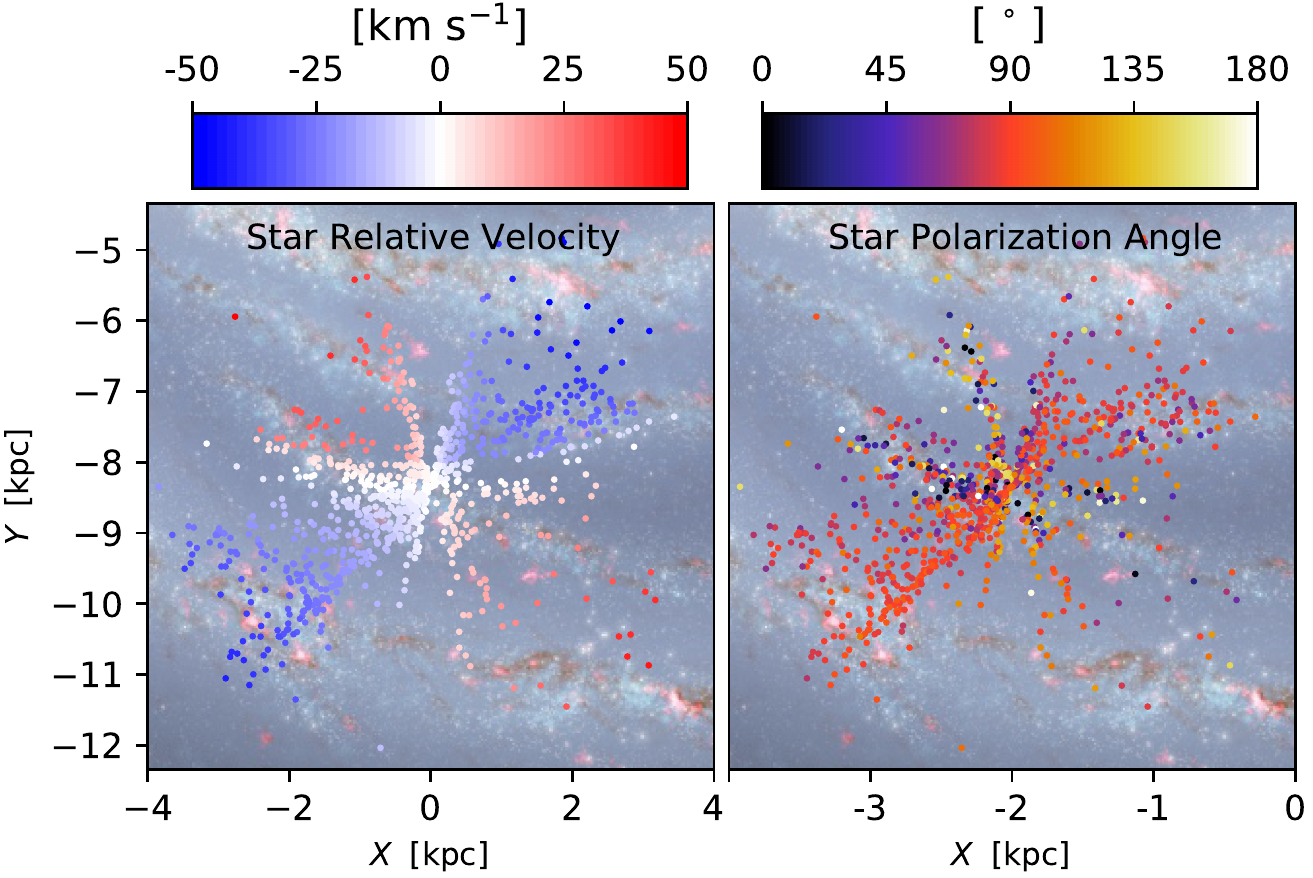}
\caption{Locations of the 1043 stars analyzed in the galaxy, projected onto the plane of the galaxy. The distance, latitude, and longitude are obtained from the \citet{heiles2000} catalog. {\it Left panel} shows the relative velocity of each star assuming only circular motions. {\it Right panel} shows the stellar polarization angle as given in the \citet{heiles2000} catalog. As a background we show the Milky Way with the Scutum-Centaurus, Sagittarius, and Perseus spiral arms along with the Orion Spur where the Sun is located.  The velocity has a visual limit of $|50|$~km~s$^{-1}$, for contrast with the Milky Way background.}
\label{chap:mapping-fig:intro}
\end{figure*}

The procedure of sub-block averaging in YL17 allows us to calculate both gradients with higher accuracy. In practical terms, the gradients are calculated similarly to VCG, but on each of the slices of the velocity/intensity channel map for RVCG and VChG respectively. Each slice is interpolated using the bicubic spline approximation over a rectangular mesh, adding 10 extra cells between each of the original points. At each of the original cells and for each slice, the direction of the gradient that indicates the magnetic field is in the direction of the cell that maximizes$\nabla^S(\mathbf{X},v_z)$ with:
\begin{equation}
\label{chap:mapping-eq:gradient}
\nabla^S(\mathbf{X},v_z) =  \frac{|S(\mathbf{X},v_z)- S(\mathbf{X}+\mathbf{X'},v_z)|}{|\mathbf{X'}|} \;,
\end{equation}
where $\mathbf{X}$ and $\mathbf{X'}$ are points in the POS, $S$ is the intensity at the thin channel that mimics the velocity or the reduced velocity centroid, and $v_z$ is the center velocity for each slice of the velocity centroid channel map. $\mathbf{X'}$ is defined in an annulus with a radius of 10 cells around $\mathbf{X}$.

The Gaussian-finding YL17 procedure enhances the accuracy of the technique compared to its original implementation of GL17 by allowing more cell (angle) possibilities to find the maximum gradient due to the interpolation process. The accuracy of the approach allows us to evaluate the precision of the magnetic field tracing.

The aforementioned procedure divides the data into blocks of cells for each of the slices. The way the YL17 procedure can evaluate the precision of the direction is by modifying the number of cells in each block. The more cells there are, the better-defined the Gaussian-like distribution--and therefore the direction of the velocity gradient--will be. However, there is a point at which adding more cells does not result in any new information and only loses spatial resolution.

For this work, we found that blocks of $21^2$ cells each (or ∼10\degree.5) give us the convergence to derive a direction inside the block of cells and preserve the spatial resolution. At each block, the most common direction of the gradient is determined and assigned to each block. This procedure was tested on the GALFA-H{\sc i} DR1 survey and the alignment was corroborated with {\it Planck} data. These block sizes are bigger than the ones needed in purely MHD simulations, due to the extra forces at play in the galactic ecosystem. These forces may even modify the gradients in such a manner that the derived direction is not indicative of the magnetic field direction.

\begin{figure*}
\centering\includegraphics[width=0.8\linewidth,clip=true]{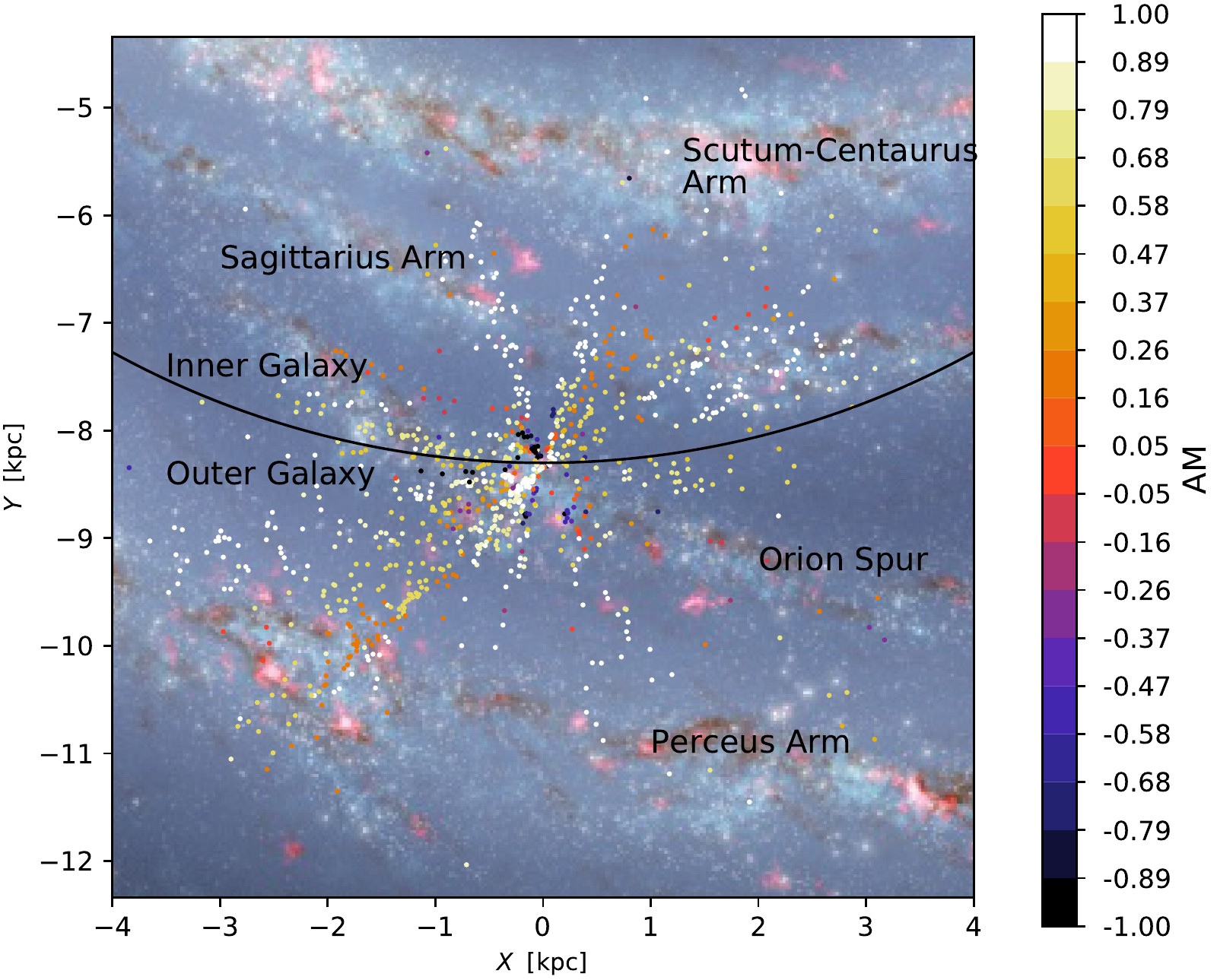}
\caption{The Alignment Measure ($AM$) for each star between the stellar polarization and the inferred magnetic field direction for their location, showing relative location in the galaxy. We used a thickness of 10~km~s$^{-1}$ for our velocity channel using the RVCG.  An $AM$ of 1 implies a parallel configuration between the two vector measurements. The solar circle is shown; anything inside it corresponds to the inner galaxy where the distance and velocity do not have a one to one correspondence. }
\label{chap:mapping-fig:AM}
\end{figure*}

\begin{figure}
\centering\includegraphics[width=0.8\linewidth,clip=true]{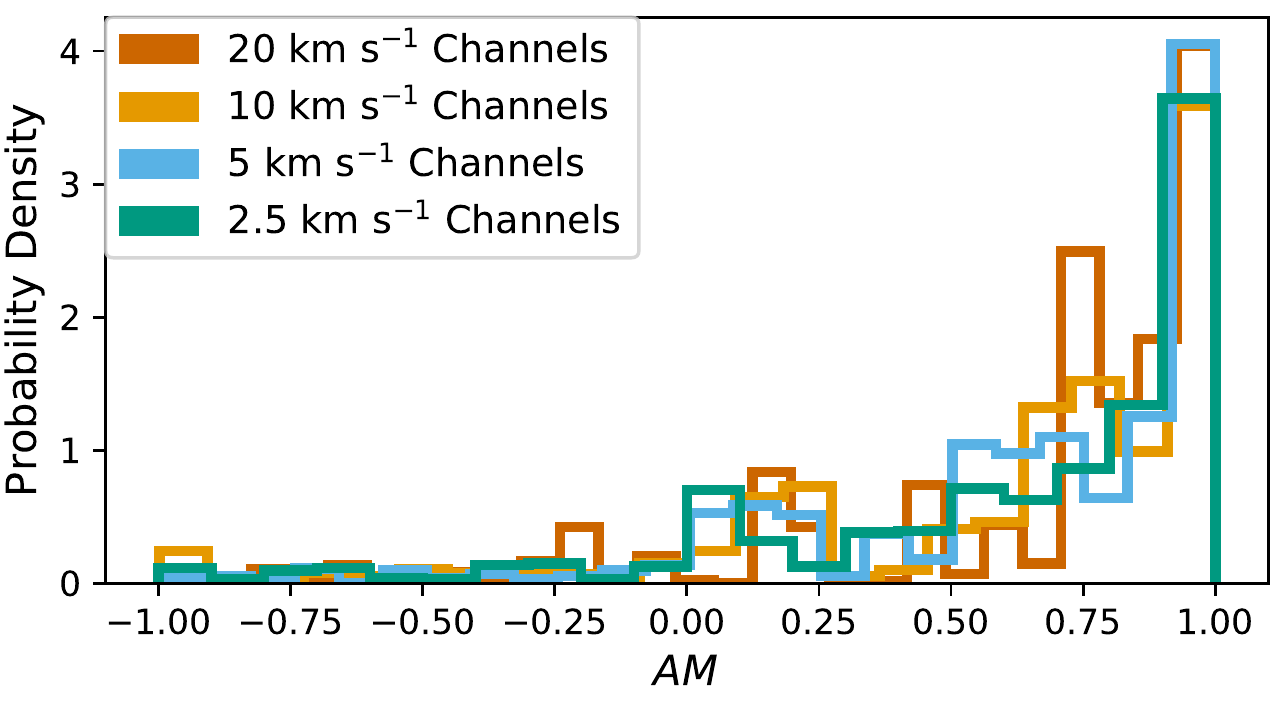}
\caption{The histogram of the Alignment Measure ($AM$) between the stellar polarization and the inferred magnetic field from the VGT. The different colors indicate the different channel widths used for the RVCG. It is clear that the alignment is similar for all the velocity thicknesses used and for all cases the mean is $\ge$0.9.}
\label{fig:CDF}
\end{figure}

For our work, we use four values (20, 10, 5, and 2.5~km~s$^{-1}$ for the channel widths when constructing the velocity channel maps for the RVCG, and 2.5~km~s$^{-1}$ for the VChG. The velocity range used is −123 to 156~km~s$^{-1}$, covering the {\it Hipparcos} instrument-limited 4~kpc radius around the Sun used for our measurements. Future measurements from {\it Gaia} will provide higher-precision estimates of the distances, which will allow us to extend this analysis to greater distances—in particular, out of the plane of the Galaxy \citep{gaia2018a, gaia2018b}.

\section{Velocity gradient analysis and the alignment with star polarization} \label{chap:mapping-sec:results}

The stellar polarization direction corresponds to the LOS integrated direction from the observer to the star. The magnetic field direction obtained with the velocity gradients corresponds to a local measurement (in velocity). The localized velocity measurement has an uncertainty arising from the turbulent velocity dispersion. To compare both measurements, we first transform the position of a star in position–position–position (PPP) to PPV\footnote{In contrast, in subsection \ref{sec:maps} we transform the results from the RVCG (that are velocity dependent) to a distance to create a first map of the POS galactic magnetic field.}.

This transformation is done assuming a galactic rotation curve given by \citet{mcclure2007}:

\begin{equation}
\label{chap:mapping-eq:rotCurve}
V = \Bigg(0.186\Big(\frac{R}{R_\odot}\Big)+0.887\Bigg) V_\odot  \;,
\end{equation}
where $R_\odot$ is the distance from the center of our Galaxy to the Sun ($R_\odot$8.34~kpc), $V_\odot$ is the circular velocity at the Sun’s location ($V_\odot$=220~km~s$^{-1}$), $V$ the circular velocity of an object (3~kpc $<$ R $<$ 8~kpc).  After 8~kpc we assume a flat rotation curve. From the star catalog we know both the spatial coordinates and its distance.  With that we can determine the velocity of the star and then the relative velocity ($V_r$) given by

\begin{equation}
\label{chap:mapping-eq:rotCurve2}
V_r = R_\odot \; \textrm{sin}(l)\Bigg(\frac{V}{R}-\frac{V_\odot}{R_\odot}\Bigg)  \;.
\end{equation}

The relative velocity is the same as the LOS velocity in the H{\sc i} PPV cube.  Figure \ref{chap:mapping-fig:intro} shows the relative velocity and polarization angle for each of the stars analyzed. Once the relative velocity is obtained, the stars are binned in different channels according to each star’s specific velocity and the thickness of the channels, to construct a PPV cube with the observed stellar polarization information. This assumes that all the stars in the stellar catalog behave consistently with the bulk properties of the Milky Way rotation.

The conversion between distances and velocities assumes a galactic rotation curve, with a 220~km~s$^{-1}$ velocity at the solar circle. As discussed in the paper \citet{mcclure2007}, a different value for the velocity would not drastically change the quality of the rotation curve. Furthermore, because we worked with relative velocities, there are no significant changes in the velocity estimations. As discussed in Section \ref{chap:mapping-sec:data}, we are not working with stars with galactic longitude around the galactic center and anticenter. This is because the velocities collapse/expand, precluding the relation of velocities to distances \citep{hartmann1999}.

With the velocity transformation, both measurements have the same spatial relation—but different information regarding the magnetic field direction. Our approach with gradients provides local measurements of the magnetic field, while the stellar polarization corresponds to an LOS integrated measurement. We therefore have to transform the results from the gradients to an LOS integrated measurement. We do this following the procedure described in\citet{clark2018}, that constructs ``pseudo" $Q$ and $U$ Stokes parameters from the directions given by density structures and then adds them up. The ``pseudo" $Q$ and $U$ Stokes are integrated along the LOS following:

\begin{equation}
\label{chap:mapping-eq:stockes2}
\phi = \frac{1}{2} \textrm{tan} \Bigg( \frac{ \int U(\mathbf{X},v_z) dv_z }{ \int Q(\mathbf{X},v_z) dv_z } \Bigg) \;,
\end{equation}
where the velocity integration corresponds the LOS integration in velocity space.  This integral could also be done in space instead of velocity. The ``pseudo" Stokes parameters are:
\begin{align}
Q(\mathbf{X},v_z) = I(\mathbf{X},v_z)\; \textrm{cos}(2\theta(\mathbf{X},v_z))\;,\nonumber \\
U(\mathbf{X},v_z) = I(\mathbf{X},v_z)\; \textrm{sin}(2\theta(\mathbf{X},v_z))\;, \label{chap:mapping-eq:stocke}
\end{align}
where $I$ is the H{\sc i} intensity at each cell in PPV, $\theta$ is the angle of the magnetic field from the VGT.

The \citet{clark2018} approach was introduced for high-latitude H{\sc i} observations, while we are using it for the disk of the Galaxy. Furthermore, \citet{clark2018} constructed the Stokes parameters from H{\sc i} filaments, while we construct them from gradients. We feel that we can use the same approach in the disk because we have a defined rotation curve that allows us to separate the contribution coming from different distances along the LOS. As noted before, we subdivide the velocity space into slices with different thickness up to the order of the turbulent velocity dispersion. With this adopted subdivision, the determination of the regions over which we add up pseudo-Stokes parameters becomes physically meaningful because it is done in physically distinct regions. In these physically distinct regions, we use velocity gradients to determine the Stokes parameters based on their velocity and density. Therefore, constructing and summing the Stokes parameters in this way builds upon the intrinsic physical properties.

The method described by \citet{clark2018} of using H{\sc i} to normalize the Stockes parameters was developed for observations above the galactic plane where there is low absorption. We use this procedure to normalize the parameters because it is known that H{\sc i} traces dust polarization (Equation \ref{chap:maping-eq:stocke}).  Since dust is what can produce a change in the direction of the polarized light it is important to take it into account \citep{andersson2015}.

For each star we now have the stellar polarization and the inferred magnetic field direction from the VGT in a format that we can compare. We compared them with the alignment measure, $AM$ (see GL17):
\begin{equation}
\label{eq:am}
AM = 2\Big\langle cos(\xi)^2-\frac{1}{2} \Big\rangle \;,
\end{equation}
where $\xi$ is the angle between the two vectors. The $AM$ is  well-suited to compare the alignment between two vectors as it precisely measures the properties of circular data (just like angle measurements). The $AM$ is widely used in dust alignment theory. The $AM$ ranges from -1 to 1, with $AM=1$ for a parallel or anti-parallel configuration and $AM=-1$ for a perpendicular one.

Figure \ref{chap:mapping-fig:AM} shows the $AM$ between the two measurements for the 10~km~s$^{-1}$ channel width using RVCG. Figure \ref{chap:mapping-fig:CDF} shows the same information in histogram form but for the four different channel widths. In general, we observe a good alignment, regardless of channel width, for most of the stars.

Toward the galactic plane, there is an abundance of molecular clouds—and therefore dust. This dust will affect the stellar polarization light, but will not be accounted for in the normalization procedure used -- because we used only H{\sc i} data. As seen in Figure \ref{chap:mapping-fig:AM}, there are regions close to the spiral arms where the AM is negative, showing an ``anti-alignment". We believe that dust from molecular clouds is what is producing this effect in the polarization. To account for that, CO emission should be incorporated.

This is the first attempt at doing such an analysis, and more studies will be necessary. With better distance measurements to stars and more star polarization measurements, we are going to address these issues. {\it Gaia} \citep{gaia2018a,gaia2018b} is the project now providing more star distances that, combined with the Galactic Plane Infrared Polarization Survey \citep{clemens2012}, will provide many more starlight polarization measurements with associated distances.

An important consideration when using the VGT is that the technique only gives the direction of the magnetic field when the magnetic field is the source of the anisotropy; this is true for sub-Alfv\'enic turbulence. In the case of super-Alfv\'enic turbulence, the difference in the magnetic field direction between polarization and gradients can be important. Super-Alfv\'enic turbulence is common in the galactic disk, and we believe that the direction we identify in our modified picture (see Figure \ref{chap:mapping-fig:AM}) can correspond to that.

It is also important to note that, for the I and IV galactic quadrants (inside the solar radius), the velocity information corresponds to two different spatial positions in the galactic disk. Therefore, using the VGT in these quadrants gives an ambiguous direction of the magnetic field that might be incorrect. As seen in Section~\ref{chap:mapping-sec:maps}, the POS magnetic field across the Galaxy does not deviate strongly, allowing us to suggest that the measurements in the I and IV quadrants should present the actual magnetic field. \citet{soler2016}, using {\it Planck}, also found that the POS magnetic field comes from a local environment similar to our findings. Even with a mildly changing magnetic field, the average AM for the inner Galaxy and for the outer one does not show measurable differences. The inner galactic AM is 0.75, while the outer part is 0.74, and the average one is 0.75 (all this pertains to the RVCG with~km~s$^{-1}$ channel width).

\begin{figure*}
\centering\includegraphics[width=0.8\linewidth,clip=true]{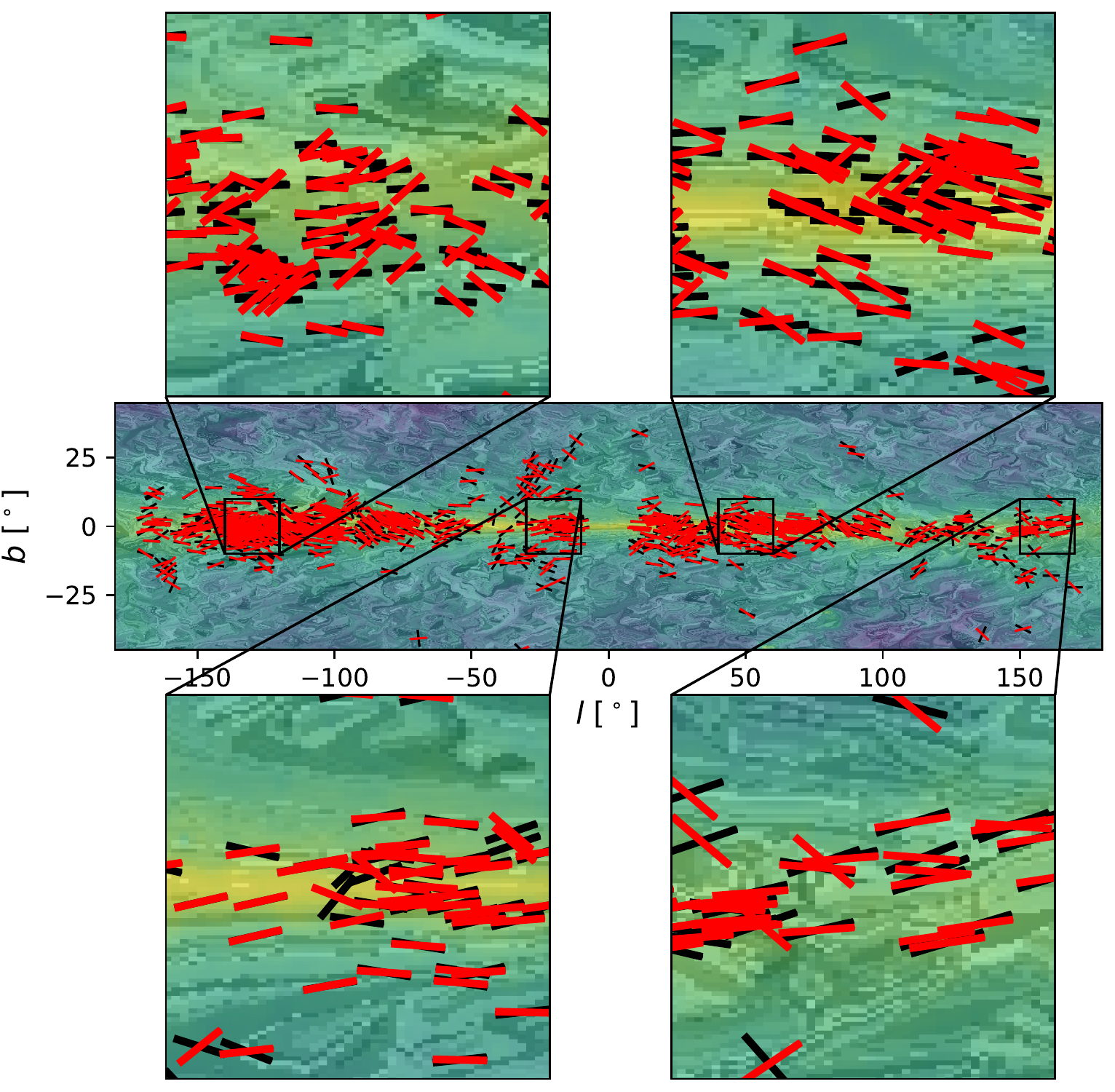}
\caption{In {\it black} the stellar polarization vectors from the \citet{heiles2000} catalog, in {\it red} the magnetic field direction obtain from the VGT. The inferred magnetic field directions from the VGT correspond to the same distance to the star, proving the same environment. The background is the H{\sc i} intensity with a line integral convolution showing the magnetic field, where there is no stellar polarization information.}
\label{chap:mapping-fig:2DMap}
\end{figure*}

Using star polarization, we have shown that the magnetic field direction inferred from the velocity gradients yields an accurate measure of the POS magnetic field in the 4 kpc solar neighborhood for which we can verify our results with starlight polarization. This suggests that our technique can be used to determine the magnetic field in regions where we do not have stellar polarization measurements (see Figure \ref{chap:mapping-fig:2DMap}). This map shows how the magnetic field in the halo is mostly perpendicular to the disk—but in the disk, it is aligned with it. This map can be compared to similar maps like those from \citet{fosalba2002} or \citet{planck2016}. It is important, however, to understand the differences in scale and resolution. Our maps have a 10\degree.5 resolution, which can exaggerate and blend features. Is also important to note that the VGT can only reliably obtain the magnetic field in velocity space (PPV), and therefore the structure of the magnetic field shown in Figure \ref{chap:mapping-fig:2DMap} may present artifacts of the assumption used to transform velocities to distances.

\section{Three-dimensional Structure of the POS Magnetic Field} \label{chap:mapping-sec:maps}

\begin{figure*}
\centering\includegraphics[width=0.8\linewidth,clip=true]{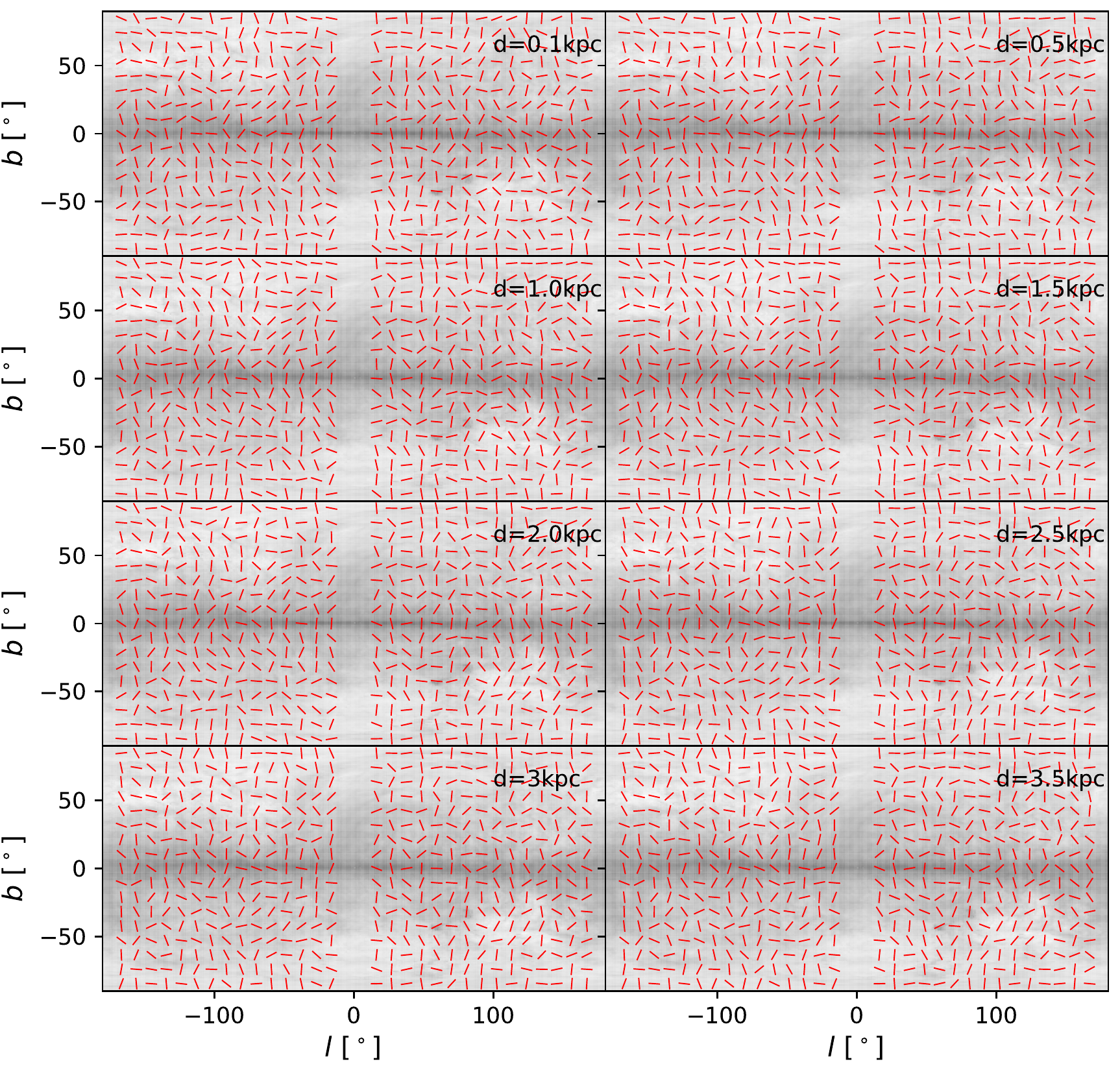}
\caption{The galactic magnetic field at eight different distances from the Sun given by the different panels. The inferred magnetic field from the VGT is shown in {\it red} with the H{\sc i} intensity map in the background. The distances are derived from the radial velocities from spectroscopic observations assuming a galactic rotation curve. Based on the rotation curve assumption, all distance estimations outside the galactic plane are not accurate.  The VGT does however find the correct information in velocity space (PPV). The erratic behaviour in the halo magnetic field is due to the assumption used and therefore is not the real structure of the halo's magnetic field. }
\label{chap:mapping-fig:DistMap}
\end{figure*}

From Section \ref{chap:mapping-sec:results} we know we can obtain the POS magnetic field in any direction as a function of the relative velocity.  Assuming the same rotation curve we can now invert the problem and obtain the POS magnetic field in any direction as a function of the distance. For this we used Equation \ref{chap:mapping-eq:rotCurve2} and \ref{chap:mapping-eq:rotCurve} in reverse.

The galactic turbulent velocity is $\sim$~10~km~s$^{-1}$, meaning there is a limitation to the precision on the distance estimates apart from the uncertainty of the galactic rotation curve. Consequently, the channels with velocity widths smaller than the turbulent velocity cannot give more precise distance estimations. Here, we present two channels with velocity width of 2.5 and 5~km~s$^{-1}$ for the RVCG, and one with velocity width of 2.5~km~s$^{-1}$ for the VChG where their distances have to be limited by the turbulent velocity. Therefore, even with high spectral resolution, there is a minimum error to the distance, just from the galactic turbulent properties.

When the information from thin channels is used (smaller than the turbulent velocity), it is important to coarsen the grading of the gradients in terms of their velocity width. Because the magnetic field is slowly variant along the LOS, a simple average suffices to arrive at a coarse measurement. This measurement is the one that can be used to determine distances for this approach. In the case that the original information is already from a coarse grading (channel widths wider than the turbulent velocity), the method can be directly applied.

Figure \ref{chap:mapping-fig:DistMap} shows the inferred POS galactic magnetic field from the VGT. The shown magnetic field direction is the LOS-integrated direction rather than a local measurement, such that it can be compared with observations. The map shown in Figure \ref{chap:mapping-fig:DistMap} is for the smallest channel width (10~km~s$^{-1}$) that would give the highest-resolution distance information.

When doing the transformation between the relative velocities and distances, we have assumed a galactic rotation curve. This was done throughout the whole Galaxy, without any remarks on the vertical height away from the plane of the Galaxy. We can therefore expect that distance estimations in the halo of the Galaxy are subject to inaccuracies in distance but not in velocity (Figure \ref{chap:mapping-fig:DistMap}). At the same time, most of the stars used in our analysis are in the galactic plane, and therefore the distance estimations and AM should be accurate.

\section{Requirements and limitations}\label{chap:mapping-sec:reqandlim}

The requirement for the use of velocity gradients is that the media should be turbulent and magnetized. Turbulence is, however, a natural state of astrophysical fluids that are characterized by high Reynolds numbers. This is the case of the galactic H{\sc i} that is turbulent and in a magnetized medium. It is important to understand that one does not need to ``believe" in the existence of turbulence in the media that we apply our technique. Studies of galactic H{\sc i} using different tools testify of the turbulent nature of the medium \citep[e.g.,][]{lazarian2000,chepurnov2010}. A similar approach can be used to justify the application of our technique to other media, such as the molecular medium. For instance, the turbulent spectra of $^{13}$CO was measured in \citet{padoan2009} using the velocity coordinate spectrum technique \citep[see more examples in][]{lazarian2009}.

The velocity gradient implementation gives the direction of the POS magnetic field with high accuracy. Most numerical studies of the technique have been done with sub-Alfv\'enic turbulence. However, results in LY18 show that low-k spatial filtering provides a way to study super-Alfv\'enic turbulence. In the presence of gravity, the VGT has to be modified to take such motions into account \citep{yuen2017b, lazarian2018b}. Additional modifications may be important in the vicinity of spiral arms, the galactic center, and high-velocity clouds.

YL17 used a statistical approach to select the peak of the distribution as the local mean magnetic field measurement. This approach provides a reliable way to study the magnetic field direction and estimate the accuracy of the direction determination. The limitation of using sub-block averaging is the decrease in map resolution. In cases where other effects are present, such as gravity, the blocks might need to be bigger—as is the case in this work. This means a decrease in the resolution of the magnetic field maps by a factor from 5 to 8 compared to the original spectroscopic maps. With high-resolution instruments available, this decrease is a reasonable price to pay for obtaining magnetic field structure. Future analysis to obtain the full sky POS magnetic field should get higher spectral resolution maps, such as the ones from GALFA or H14PI \citep{galfa, hi4pi}.

Our present study provides the 3D distribution of POS magnetic field. As we discussed, the accuracy of obtaining the distance along the LOS depends on the profile determined by the rotation curve in the given direction. The coarse grading is determined by the uncertainty associated with turbulent dispersion. There are several problems associated with a rotation curve, the most important of which is that, depending on the rotation curve used, the distance estimations will be different. We use the\citet{mcclure2007} rotation curve that is known to deviate from other rotation curves like the ``universal rotation curve'' of \citet{reid2014}. Because our understanding of the galaxyʼs rotation and its topology will continue to evolve, it is important to remember that the reported velocities are model-independent and will always be truthful. Furthermore, we know that, even in the galactic disk, there are motions not covered in a rotation curve that can and should be taken into account when transforming velocities to distances \citep{tchernyshyov2017}.

\section{Discussion and Outlook}\label{chap:mapping-sec:discussion} 

\subsection{Two- and Three-dimensional Information Available with Velocity Gradients}

All earlier attempts to employ velocity gradients have been focused on obtaining the POS information of the magnetic field distribution \citep[see][LY18]{casanova2017b}. The limitations of this approach stem from the fact that the LOS additions of the velocity gradient are different from the Stokes parameters, and this limits the extent to which the directions obtained with polarization and magnetic field can be compared.

More importantly, it is very valuable to obtain 3D information regarding the distribution of the POS magnetic field. The rough map of 3D distributions of H{\sc i} in space is available using the galactic rotation curve. Using our new tools, i.e., VChGs and RVCGs (LY18) we are able to study POS magnetic field distribution as a function of the distance along the LOS.

To test the accuracy of the new way of studying POS magnetic field 3D distribution, we used the polarization from stars to which the distances were known. These stars were used as beacons to probe the accuracy of our 3D maps along the given directions. To make the corresponding comparison, we used the procedure of constructing of ``pseudo-Stokes" parameters as in \citet{clark2018}. Using the stars within 4~kpc of the Sun, we obtained a good correspondence between the polarization expected on the basis of our 3D maps and the actual starlight polarization. This gives us confidence that velocity gradients are able to study the 3D distribution of the POS magnetic fields in the Milky Way at arbitrary distances from the observer.

We expect that knowledge of the 3D distribution of the galactic magnetic field can help to understand key astrophysical processes, including the processes of star formation, propaga- tion and acceleration of cosmic rays, etc. For instance, the presence of magnetic fields plays a critical role in the process of star formation, because magnetic fields add an additional pressure acting against gravity \citep[e.g.,][]{mckee2007}. The knowledge of the 3D structure of the POS magnetic field can significantly help in understanding the nature of this type of support.

Detailed information about magnetic fields is also essential for comparing observations and MHD galactic simulations \citep[e.g.,][]{shetty2006, dobbs2008}. The 3D information that is available with velocity gradients provides constraints on the morphology of the field in numerical simulations, affecting our understanding of the transport of angular momentum as well as pressure support in our own Galaxy \citep{kim2017}.

\subsection{Studies of CMB polarization}

Our study also provides a way to help in the search for cosmological B-mode polarization. The corresponding polarized signal arising from these modes is orders of magnitude less than the polarized signal from the foreground-aligned galactic dust. These B-modes, if detected, would be definitive proof for inflation \citep{seljak1997}.

This work presents the first measurement of the POS magnetic field embedded in the Milky Way, presenting its direction as a function of the distance from us (for the case of the disk and as a function of velocity for the halo). With both ``local'' (up to the turbulent velocity) and integrated measurements along the LOS, one can study the magnetic field in a similar fashion to the {\it Planck} mission.

The {\it Planck} mission aimed to understand the cosmological properties of our universe, but realized that it was fundamental to first fully understand the local galactic polarization environment to accurately derive cosmological parameters \citep{Planck2016a, Planck2016b}. Our map, while lower in spatial resolution, does give a 3D component of the magnetic field. This information should help constrain galactic parameters such as the polarization fraction and the magnetization that eventually will help constrain cosmological properties.

Spatial regions for CMB polarization analysis can be found where the magnetic field does not change drastically along the LOS. This removes the contribution of the Milky Way to the measurements. The search can be done using the data from the VGT in velocity space. Once these regions are detected, one can use optical starlight polarization measurements with parallax to better constrain the properties of the CMB \citep{tassis2015}.

This direction of research is similar to the one in \citet{clark2018}, with the important distinction that we can use velocity gradients and not filaments. Elsewhere, we plan to compare our technique with \citet{clark2018}.

\subsection{3D distribution of media magnetization}

Far-infrared polarization measurements are a common way to obtain the magnetic field direction and infer the magnetic field strength. The shortcoming of these measurements, however, is that they suffer from instrumental polarization and frequently require expensive space or balloon-borne missions. Uncertainties of the grain alignment theory \citep[see][]{lazarian2007}, as well as the failure of grain alignment to trace magnetic fields at high optical depths \citep{andersson2015}, provide limitations for measuring magnetic fields with polarimetry.

A new way of obtaining the Alfv\'en Mach number $M_A=V_L/V_A$, where $V_L$ is the turbulent velocity and $V_A$ is the Alfv\'en velocity is proposed in \citet{lazarian2018}. The technique is based on the fact that the distribution of velocity gradients measured within the sub-block of data points employed for the sub-block averaging depends on $M_A$. Within the approach that we employed in this paper, the 3D distribution of $M_A$ should be obtained.

Frequently, the estimated strength of the magnetic field is derived by applying the Chandrasekhar–Fermi (C-F) method to polarization measurements \citep{chandrasekhar1953}. Our approach based on velocity gradients offers a novel alternative. It was noted in GL17 that the strength of the magnetic field can also be derived using velocity gradients with a method analogous to the C-F method or via an independently derived method \citep[GL17]{lazarian2018}. The major difference is the use of the dispersion of the observed VCG instead of the dispersion of the dust polarization measurements. The estimation of the magnetic field strength has also been expanded to media with self-absorption, e.g., molecular emission from CO \citep{casanova2017b}. In light of our present study, it means that the 3D distribution of magnetic field strengths can be obtained for the galactic H{\sc i}.

\subsection{Expanding the range of applicability of velocity gradients}

The VGT is a rapidly developing field. For our studies, we employed H I, for which the effect of self-absorption is not important. For $^{13}$CO -- and especially for $^{12}$CO -- the self-absorption is an essential effect. Nevertheless, it is shown in \citep{casanova2017} that the VCGs can be successfully used in such media to trace magnetic fields. A similar conclusion was also reached in a subsequent study in \citet{hsieh2018}. In fact, 3D tomography of magnetic fields in molecular clouds is possible through combining the emission lines that originate at different depths within molecular clouds.

A comparison of the velocity gradients with other techniques for finding the magnetic field directions using the turbulence anisotropy has shown the advantages of velocity gradients. For instance, it was shown in \citet{yuen2017b} that the velocity gradients provide a more detailed tracing of magnetic field compared to the correlation function anisotropy (CFA) technique \citep{lazarian2002, esquivel2005}. At the same time, it may be advantageous to combine both techniques, as the anisotropies revealed by the CFA technique can be used to distinguish the compressible and incompressible contributions to the interstellar turbulence \citep{kandel2016a, kandel2016b}.

\subsection{Complementary Ways to Probe the 3D Magnetic Field
Structure}

We have mentioned already that the starlight polarization provides a way to probe the 3D magnetic field structure along the directions toward stars. In this paper, this fact was used to test our predictions of the galactic 3D magnetic field distribution.

Another promising way is related to the use of the synchrotron polarization gradient (SPGs) \citep{lazarian2018b}. Magnetic turbulence induces synchrotron fluctuations-- the statistics of which has been recently described in \citet{lazarian2012b,lazarian2016}. The synchrotron fluctuations reflect the magnetic fluctuations. For Alfv\'enic turbulence, the fluctuations of magnetic field and velocity enter symmetrically. Therefore, it is clear that the synchrotron gradients, e.g., synchrotron intensity gradients (SIGs), are expected to trace magnetic fields similarly to the way velocity gradients trace magnetic field. The SIG technique was introduced in \citet{lazarian2017}. Similarly, it was demonstrated, in \citet{lazarian2018b} that the SPGs also successfully trace interstellar magnetic fields. The advantage of SPGs compared to SIGs, however, is that, due to the Faraday depolarization at low frequencies, the SPGs trace magnetic fields up to a certain distance along the LOS. By changing this frequency, \citet{lazarian2018b} demonstrated the possibility of mapping the 3D distribution of the magnetic field via synthetic observations.

In terms of the present study, the possibility of having 3D maps obtained with the synchrotron technique is very attractive. It is important to understand that, in general, H I and synchrotron emission probe different phases of the ISM; \citet[see][for the list of the ISM phases]{draine2011}. For many astrophysical problems, e.g., cosmic ray propagation \citep[see][]{schlickeiser2007} and star formation \citep[see][]{mckee2007}, it is important to know how magnetic fields in different ISM phases are interrelated.

Naturally, the velocity gradients are complimentary to other techniques. At the beginning of the paper, we mentioned the Zeeman splitting technique. Obtaining the LOS value of the magnetic field is very advantageous and it can be used together with measurements of the plane-of-sky magnetization that is available through studies of the dispersion of the velocity gradient distribution \citep{lazarian2018} or the Chandrasekhar–Fermi approach based on gradients \citep{casanova2017, casanova2017b} to find the 3D vector of magnetic field. Similarly, the VGT can be used together with a new, very promising way of studying magnetic fields with the ground state atomic alignment; \citet[see][for a review]{yan2015}. The latter technique uses the polarization of spectral lines, and therefore the same spectral line data can be synergistically used to trace magnetic field.

\section{Conclusion}\label{chap:mapping-sec:conclusion}

In this work, we have constructed the first map of the galactic POS magnetic field using two approaches that make use of Velocity Gradient Technique: namely, VChGs and RVCGs. The velocity gradients were applied to the H I LAB survey. Both techniques allow the measurement of the POS magnetic field as a function of the relative velocity. Using a galactic rotation curve, we transformed the relative velocity measurements to distance estimations (in the galactic disk). With these, we were able to compare the inferred magnetic field direction from the VGT to the magnetic field obtain from stellar polarization. We used 1043 star polarization measurement from the Heiles (2000) catalog to test our 3D magnetic field maps. The alignment between the two measurements was quantified, and our results testify that our derived 3D POS galactic magnetic field map provides an accurate description of the Galaxyʼs actual magnetic field distribution.

\acknowledgments

We thank Julie Davis, Zach Pace, and Ka Ho Yuen for the insightful discussions. Travel support by NSF ACI 1713782 is acknowledged. Partial support for DFGC was provided by CONACyT (Mexico).

\bibliographystyle{apj}
\bibliography{biblio-gradobs}{}

\end{document}